%% file: main.tex
\documentclass[a4paper,11pt]{article}

\input{auxiliary/packages_and_formatting}

\title{\boldmath Treating Detector Systematics via a Likelihood Free Inference Method}

\author[a]{L. Fischer \orcidlink{0000-0002-7645-8048},}
\author[a]{R. Naab \orcidlink{0000-0003-2512-466X},}
\author[a, b]{and A. Trettin \orcidlink{0000-0003-0350-3597}}

\affiliation[a]{Deutsches Elektronen-Synchrotron DESY, Platanenallee 6, 15738 Zeuthen, Germany}
\affiliation[b]{University of Manchester, M13 9PL Manchester, UK}

\emailAdd{leander.fischer@desy.de, richard.naab@desy.de, alexandra.trettin@manchester.ac.uk}

\input{sections/abstract}


\arxivnumber{2305.02257}

\begin{document}
\maketitle
\flushbottom

\input{sections/introduction}
\input{sections/motivation}

\input{sections/classification_methods}
\input{sections/toy_mc}

\input{sections/conclusion}

\input{sections/acknowledgements}

\bibliography{auxiliary/MyBibFile}
\bibliographystyle{JHEP}


\appendix
\input{sections/supporting}

\end{document}

%% file: auxiliary/packages_and_formatting.tex
\usepackage{jinstpub}

\usepackage[utf8]{inputenc}
\usepackage[ngerman,english]{babel}

\usepackage{url}
\usepackage{siunitx}
\usepackage{physics}
\usepackage{graphicx}
\usepackage[
    noabbrev,
    capitalise,
    nameinlink,
]{cleveref}

\usepackage{orcidlink}

\usepackage{pgfplots}
\usepackage{pgfplotstable}
\ifluatex
\else
    \DeclareUnicodeCharacter{2212}{−}
\fi
\usepgfplotslibrary{groupplots,dateplot}
\usetikzlibrary{patterns,shapes,arrows, decorations.markings}
\pgfplotsset{compat=newest}
\usepgfplotslibrary{fillbetween}

\definecolor{black}{RGB}{0,0,0}
\definecolor{orange}{RGB}{230, 159, 0}
\definecolor{skyblue}{RGB}{86, 180, 233}

\definecolor{tab_orange}{RGB}{255, 128, 14}
\definecolor{tab_blue}{RGB}{0, 107, 164}

\definecolor{bluishgreen}{RGB}{0, 158, 115}
\definecolor{yellow}{RGB}{240, 228, 66}
\definecolor{blue}{RGB}{0, 114, 178}
\definecolor{vermilion}{RGB}{213, 94, 0}
\definecolor{reddishpurple}{RGB}{204, 121, 167}

\definecolor{lightgray204}{RGB}{204,204,204}

\newcommand{\bs}{\boldsymbol}

%% file: sections/abstract.tex
\abstract{
Estimating the impact of systematic uncertainties in particle physics experiments is challenging, especially since the detector response is unknown analytically in most situations and needs to be estimated through \textit{Monte Carlo} (MC) simulations. Typically, detector property variations are parameterized in ways that implicitly assume a specific physics model, which can introduce biases on quantities measured by an analysis. In this paper, we present a method to recover a model-independent, event-wise estimation of the detector response variation by applying a likelihood-free inference method to a set of MC simulations representing discrete detector realizations. The method provides a re-weighting scheme for every event, which can be used to apply the effects of detector property variations fully decoupled from the assumed physics model.
Using a toy MC example inspired by fixed-baseline neutrino oscillation experiments, we demonstrate the performance of our method. We show that it fully decouples the modeling of the detector response from the physics parameters to be measured in a MC forward-folding analysis.
}

%% file: sections/introduction.tex
\section{Introduction} \label{sec:intro}

The \textit{Standard Model} (SM) of particle physics has been tested to remarkable precision and seemingly withstands all attempts to disprove it. However, there are some phenomena that it cannot explain, such as non-zero neutrino masses. In order to uncover the nature of these phenomena, the field needs to move towards an era of precision measurements using large amounts of data. As the statistical power of these analyses increases, the need for precise treatments of systematic uncertainties becomes a growing concern. Estimating the impact of systematic uncertainties related to imperfect detector knowledge poses a particular challenge in many experiments, as the detector response is typically unknown analytically and must be approximated using MC simulations. Deriving the effect of detector response uncertainty necessitates running the simulation under various detector property assumptions, resulting in multiple sets of simulations with different detector realizations. One approach to account for the resulting variation regarding a changing detector response is to re-evaluate observable distributions from different simulation sets and parameterize the corresponding shift in the final analysis binning \cite{OVS_PRD}. However, to get the observable distributions, an explicit assumption of the physics parameters has to be made, which intrinsically couples the parameterized variations to the assumed physics model.

Here, we present a method for obtaining a model-independent, event-wise estimate of detector response variation via a likelihood-free inference approach, fully decoupling the effects of detector property variations from the assumed physics model. We focus on neutrino oscillations \cite{neutrino_oscillationso_og} as an example to showcase the method's effectiveness. Neutrino oscillations are a field of precision measurements, where many \textit{Beyond Standard Model} (BSM) theories have been proposed to explain the experimental observations and testing those models requires handling of large statistics datasets with accurate modeling of detector uncertainties. Our method can be applied in numerous cases and offers the advantage of more efficient use of MC simulations, reducing the computational resources required for MC production, and hence saving power and reducing CO2 emissions. Additionally, our method is binning independent, allowing for efficient analysis optimization studies and going beyond traditional binned likelihood method analysis. We anticipate that this flexibility will enable more accurate and comprehensive analyses in a wide range of high-energy physics experiments.

In \cref{sec:systematics_via_llh_free_inference} we formulate the problem of the classical approach in detail and explain the likelihood free approach, which is then applied to a simplistic toy MC using a k-nearest neighbor (KNN) classifier in \cref{sec:method-description}. In \cref{sec:toy-mc} the method is applied to the realistic example of neutrino oscillations and the performance is estimated. Finally, we conclude in \cref{sec:conclusion} and give a short outlook of how this method can be improved further.

%% file: sections/motivation.tex
\section{Modeling Variations in Detector Response via Likelihood-Free Inference}
\label{sec:systematics_via_llh_free_inference}

\subsection{Formulation of the Problem}
\label{sec:problem-formulation}

In a MC \textit{forward-folding} analysis, physics parameters, $\boldsymbol{\theta}$, are estimated by comparing the distribution of observables measured in an experiment, $\vb{y}$, to the distribution of MC simulated events that are weighted according to the theoretical expectation given any particular value of $\boldsymbol{\theta}$. The observables $\vb{y}$ are typically reconstructed quantities, such as the energy of an event or a classification score. The MC events are drawn from a distribution of true particle properties $\Phi_{\mathrm{sim}}(\vb{x})$, where $\vb{x}$ may contain variables such as the particle energy and type.
The distribution $\Phi_{\mathrm{sim}}(\vb{x})$ must not be a physical model, but can rather be any distribution chosen such that the MC sampling covers the entire range of particle properties that may be of interest for the given analysis.
To reproduce the event distribution for a physical model, these MC events are weighted by the ratio $w_\mathrm{flux}=\Phi(\vb{x}|\bs{\theta})/\Phi_{\mathrm{sim}}(\vb{x})$, where $\Phi(\vb{x}|\bs{\theta})$ is the differential particle flux expected at the physics parameters $\boldsymbol{\theta}$. The result is referred to as \textit{re-weighted} MC.
We assume that a closed-form expression for $\Phi_{\mathrm{sim}}(\vb{x})$ and $\Phi(\vb{x}|\bs{\theta})$ exists, or at least an analytic approximation thereof, such as a spline.
Given the true properties of a particle, $\vb{x}$, the detector detects the particle with probability $P(\mathrm{acc} | \vb{x},\bs{\alpha})$ and, if the event has been detected, produces observables according to the conditional probability distribution $P(\vb{y} | \vb{x},\bs{\alpha})$. The product of the two is what we refer to as \textit{detector response}. Here, $\bs{\alpha}$ is a set of detector properties that influence the detector response and are nuisance parameters to the analysis. The detector properties are usually only known with a given uncertainty, where the expectation value is called \textit{nominal} and variations around the nominal value are referred to as \textit{off-nominal}. The comparison between data and simulation is typically made by dividing data and MC events in a histogram of the observed quantities $\vb{y}$ and then calculating the Poisson or $\chi^2$ likelihood between the data and the MC expectation in each bin. 

If the detector response was known analytically, then the expectation value, $\mu_i$, for the event count in a bin with index $i$ could be calculated by integrating out the distribution of the true particle properties, such that
\begin{equation}
    \mu_i(\boldsymbol{\theta}) = \int_{\vb{y}\in \mathrm{bin\,}i} \dd\vb{y}
    \int \dd \vb{x} P(\vb{y}|\vb{x},\bs{\alpha}) P(\mathrm{acc}|\vb{x},\bs{\alpha}) \frac{\Phi(\vb{x}|\bs{\theta})}{\Phi_{\mathrm{sim}}(\vb{x})}
    \;,\label{eq:expectation-bin-integral}
\end{equation}
where $P(\mathrm{acc} | \vb{x},\bs{\alpha})$ and $P(\vb{y} | \vb{x},\bs{\alpha})$ are the above introduced detector acceptance probability and the observable probability distribution, respectively, and $\Phi(\vb{x}|\bs{\theta})/\Phi_{\mathrm{sim}}(\vb{x}) = w_\mathrm{flux}$ is the re-weighting factor to account for the difference in sampling distribution and probed physics distribution.
In practice, there is usually no analytical expression for the detector response so that a MC simulation is performed to evaluate its effects. The expectation value is then estimated by counting the weighted MC samples such that 
 \begin{equation}
     \hat{\mu}_i(\boldsymbol{\theta}) = \sum_{j} I(\vb{y}_j \in \mathrm{bin}\,i)
     \frac{\Phi(\vb{x}_j|\bs{\theta})}{\Phi_{\mathrm{sim}}(\vb{x}_j)}
     \;,\label{eq:weighted-expectation}
 \end{equation}
 where the index $j$ runs over the MC events and $I$ is the indicator function that is equal to one if the condition inside the parentheses is fulfilled. This converges to \cref{eq:expectation-bin-integral} in the case of infinite statistics as the MC events samples are drawn from $P(\mathrm{acc} | \vb{x},\bs{\alpha}) \, P(\vb{y} | \vb{x},\bs{\alpha})$.
 
 The estimate in \cref{eq:weighted-expectation}, however, is only valid for the particular detector parameter values that were assumed during the MC generation.
 The method proposed in previous works \cite{2019_Snowstorm, OVS_PRD} to estimate $\mu_i$ as a function of the detector parameters is to find gradients $\grad_\alpha \mu_i$ of the estimate with respect to these parameters in every analysis bin.
 These gradients, however, have to be calculated for a particular choice of physics parameters because the flux $\Phi(\vb{x}|\boldsymbol{\theta})$ is integrated inside a product with the detector response over the true event parameters $\vb{x}$ as shown in \cref{eq:expectation-bin-integral}.
Therefore, gradients that are estimated using only reconstructed quantities have to be recalculated for every value of the physics parameters that are evaluated during the analysis, which may become infeasible when the number of nuisance parameters is large.

In this work, we propose a method that fully decouples the detector response from the physics parameters by finding an \emph{event-wise} parameterization for $P(\vb{y}|\vb{x},\bs{\alpha})$ and $P(\mathrm{acc}|\vb{x},\bs{\alpha})$ that depends only on the true and reconstructed quantities of each MC event.
Given this parameterization, the expectation value in each analysis bin for an arbitrary value of $\alpha$ can be calculated via re-weighting of the nominal MC set that has been generated at the detector parameters $\bs{\alpha}_\mathrm{nom}$ as
\begin{equation}
     \hat{\mu}_i(\boldsymbol{\theta}, \bs{\alpha}) = \sum_{j} I(\vb{y}_j \in \mathrm{bin}\,i)
     \frac{P(\vb{y}_j|\vb{x}_j,\bs{\alpha})P(\mathrm{acc}|\vb{x}_j,\bs{\alpha})}
     {P(\vb{y}_j|\vb{x}_j,\bs{\alpha}_\mathrm{nom})P(\mathrm{acc}|\vb{x}_j,\bs{\alpha}_\mathrm{nom})}
     \frac{\Phi(\vb{x}_j|\bs{\theta})}{\Phi_{\mathrm{sim}}(\vb{x}_j)}
     \;.\label{eq:weighted-expectation-detsys}
\end{equation}
Crucially for the method presented in this paper, the ratio 
\begin{equation}
    r_{j}(\bs{\alpha}) = \frac{P(\vb{y}_j|\vb{x}_j,\bs{\alpha})P(\mathrm{acc}|\vb{x}_j,\bs{\alpha})}
     {P(\vb{y}_j|\vb{x}_j,\bs{\alpha}_\mathrm{nom})P(\mathrm{acc}|\vb{x}_j,\bs{\alpha}_\mathrm{nom})}\label{eq:response-ratio}
\end{equation}
is fully decoupled from the physics parameters $\bs{\theta}$. The intuition for this is that the detector only reacts to the final state of any given particle, which does not depend on the overall distribution of the entire sample. However, in contrast to the flux model $\Phi(\vb{x} | \bs{\theta})$, there is usually no closed-form expression known to predict the detector response. In the following section, we will introduce a way to estimate the ratio from MC simulation instead.

\subsection{Likelihood Free Approach}

The aim of our method is to derive an approximation for the detector response ratio $r_{j}(\bs{\alpha})$, given in \cref{eq:response-ratio}, without the need to re-simulate every single event many times, which would be prohibitively expensive in terms of computation. Instead, we use a set of MC sets, off-nominal sets or systematic sets in the following, that are simulated under alternative detector response assumptions, varying the detector properties inside their uncertainty range. In the following, $k$ is the index of each MC set and $\bs{\alpha}_k$ are the detector parameters under which it was created. We furthermore denote the detector parameters of the nominal set as $\bs{\alpha}_\mathrm{nom}$.

Before we find the response ratio $r_{j}(\bs{\alpha})$ as a smooth function in $\bs{\alpha}$, we first calculate the discrete ratios $r_{jk}=r_j(\bs{\alpha}_k)$ for each event $j$ in the nominal MC set.
To achieve this, we apply Bayes' theorem \cite{sep-bayes-theorem} to \cref{eq:response-ratio} to express $r_{jk}$ as the ratio of the posterior probability that the event $j$ given its true and reconstructed properties $\vb{x}_j$ and $\vb{y}_j$, belongs to the MC set $k$
\begin{equation}
    r_{jk} = \frac{P(\vb{y}_j|\vb{x}_j,\bs{\alpha}_k)P(\mathrm{acc}|\vb{x}_j,\bs{\alpha}_k)}
     {P(\vb{y}_j|\vb{x}_j,\bs{\alpha}_\mathrm{nom})P(\mathrm{acc}|\vb{x}_j,\bs{\alpha}_\mathrm{nom})}
     =
     \frac{P(\bs{\alpha}_k|\vb{x}_j,\vb{y}_j)}{P(\bs{\alpha}_\mathrm{nom}|\vb{x}_j,\vb{y}_j)}\;. \label{eq:ratio-posterior}
\end{equation}
This relationship holds under the assumption that the initial MC sample before the application of the detector response has been drawn from the same distribution of true event properties. A detailed derivation of this relationship can be found in \cref{sec:posterior-ratio-derivation}.
The posterior distributions, $P(\bs{\alpha}_k|\vb{x}_j,\vb{y}_j)$, can be acquired from a classifier that estimates the posterior probability for an event with parameters $\vb{x}_j,\vb{y}_j$ to belong to the MC set $k$. This means that we can translate the task of finding the re-weighting factors into a \emph{classification task}. Such an inference method, where probability distributions are learned as a ratio of posteriors from a classifier, is also known as a \emph{likelihood-free inference} method. Such methods are applied in different fields of statistical inference such as event reconstructions \cite{FreeDOM}.

%% file: sections/classification_methods.tex
\textbf{}\section{Classification}
\label{sec:method-description}
In principle, any classifier that produces well-calibrated class posterior probabilities can be used to compute the posterior ratio in \cref{eq:ratio-posterior}. In this work, we are using a k-nearest neighbor (KNN) classifier \cite{knn_classifier}. Another method that could be used for this purpose is a neural network, which would also produce posterior probabilities when combined with a cross-entropy classification loss \cite{NNPosteriors}. We chose the KNN primarily for its simplicity and reliability. In contrast to neural networks that can potentially involve the tuning of dozens of hyper-parameters of the network architecture and training procedure, the KNN only has a single hyper-parameter and does not require any training. Furthermore, the output of the KNN is straight-forward to interpret and will not behave erratically towards the edges of the distribution of the input data. 
In contrast to other classification methods such as neural networks or decision trees, the KNN does not build a generalized internal model of the input distributions to compute its output score, and always requires access to the entire training data set to function. While this may be an untenable drawback for other applications such as computer vision, it is perfectly acceptable for our purposes.
Since we found that the KNN prediction is biased in places where the distributions have large gradients we modified it with the linear tilt correction described in \cref{sec:linear-tilt-correction} to remove the bias. 
We will demonstrate the performance of the classifier and the re-weighting scheme on a realistic toy example in \cref{sec:toy-mc}.

\subsection{K-Nearest Neighbors Classification}
\label{sec:simple-knn}

The KNN classifier is a non-parametric method for classification. It is widely used due to its simplicity and effectiveness, particularly in situations where the decision boundary between classes is highly non-linear.
Given a set of training samples with features $X$ and class labels $Y$, the KNN classifier assigns a new input sample $x$ to one of the available classes by identifying the $N$ samples in $X$ that are closest to $x$ in terms of some distance metric $d(\cdot,\cdot)$.
The KNN classifier then estimates the posterior probability of class $k$ for input $x$ by counting the fraction of samples in the neighborhood that belong to class $k$ as
\begin{equation}
    P(Y=k|X=x) = \frac{1}{N} \sum_{j\in \mathcal{N}_k(x)} 1\;,\label{eq:knn-posterior}
\end{equation}
where $\mathcal{N}_k(x)$ is the set of indices of the $N$ nearest neighbors of $x$ that belong to class $k$.
The only hyper-parameter of this method is the number of neighbors, $N$. A value of $N$ that is too small will lead to overfitting, as the distribution of the few samples in a neighborhood is dominated by random fluctuations. A higher value of $N$ reduces the variance by integrating over a larger number of samples, but can lead to underfitting, if the size of the neighborhood becomes larger than the typical feature size of the class boundaries. A higher value of $N$ additionally increases the computational time.

\begin{figure}[ht]
    \centering
    \includegraphics{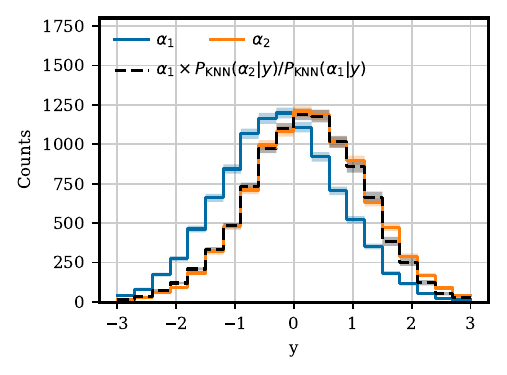}
    \caption{Simplistic toy MC example of two shifted normal distributions (blue/orange), mimicking final level observable distribution $y$ under different detector systematic variations $\alpha_1$ and $\alpha_2$, compared to a distribution obtained by re-weighting the events of the distribution related to $\alpha_1$ by the ratio of the posterior distributions predicted by a KNN classifier. As expected, the re-weighted distribution related to $\alpha_1$ matches the distribution related to $\alpha_2$ within the statistical uncertainty ($\chi^2/\mathrm{DOF}=1.85$).}
    \label{fig:simple-reweighting-illustration}
\end{figure}

\begin{figure}[ht]
    \centering\includegraphics{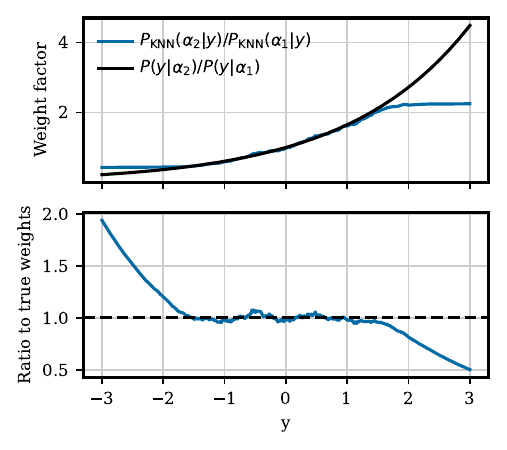}
    \caption{KNN estimated posterior ratio (blue) compared to the true analytical likelihood ratio (black) for the simplistic toy example of two shifted normal distributions.}
    \label{fig:knn-weights-gaussian-shift}
\end{figure}

To showcase how we use the KNN to estimate the posterior ratio from \cref{eq:ratio-posterior} we look at the example of a single observable distribution $y$ for two choices of a single detector systematic parameter $\alpha_1$ and $\alpha_2$. For simplicity we assume that the observable follows a normal distribution and that the detector response effect of changing from $\alpha_1$ to $\alpha_2$ is just a shift of its mean as shown by the blue/orange distributions in \cref{fig:simple-reweighting-illustration}. We assume that the acceptance is 100\% and therefore the true variable distributions are identical and don't have to be taken into account. Using a KNN with the number of neighbors set to $N=3000$ we can predict the posterior distributions $P_\mathrm{KNN}({\alpha}_1|y)$ and $P_\mathrm{KNN}({\alpha}_2|y)$ to calculate the KNN prediction of the re-weighting ratio from \cref{eq:ratio-posterior}. With this we can re-weight the blue distribution to recover the orange distribution within statistical uncertainty ($\chi^2/\mathrm{DOF}=1.85$). Since we know the analytically correct expressions for the likelihood distributions $P(y|{\alpha}_1)$ and $P(y|{\alpha}_2)$ in this example case, we can compare the KNN predicted ratio to the true, which is shown in \cref{fig:knn-weights-gaussian-shift}. The black line in the plot shows the true analytical likelihood ratio, which is expected to match the posterior ratio. The posterior ratio estimated by the KNN is close to the analytical solution within the range of $[-1.5,1.5]$, but begins to deviate due to low statistics when reaching the edges of the distributions.

\subsection{Interpolation}
\label{sec:ultrasurface-interpolation}

\begin{figure}[ht]
    \centering
    \includegraphics{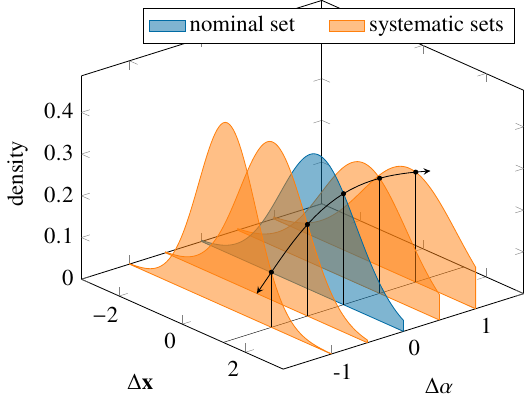}
    \caption{Illustration of the re-weighting process used to model changes in the detector response. The axis labeled $\Delta \vb{x}$ stands in for the parameters that characterize each individual event. The axis labeled $\Delta {\alpha}$ stands in for the detector parameters that vary between systematic sets. The distribution of event parameters, $\vb{x}$, are shown for the nominal MC set in blue, the off-nominal sets in yellow. The black line shows the function along which an event at a particular $\vb{x}$ is re-weighted as a function of ${\alpha}$.\label{fig:ultrasurface-illustration}}
\end{figure}

After applying the classification described in \cref{sec:simple-knn}, we can use the output of the classifier to re-weight each event $j$ in the nominal MC set to emulate any systematic MC set $k$ using the ratio $r_{jk}$ from \cref{eq:ratio-posterior}. However, for a practical data  analysis, we need the ability to smoothly interpolate between the discrete values of $\alpha_k$ that correspond to the different discrete simulation sets, meaning that we want the ratio $r_{jk}$ for an arbitrary value of $\alpha$ as in
\begin{equation}
    r_{j}(\bs{\alpha}) = \frac{{P_\mathrm{KNN}}(\bs{\alpha} | \vb{x}_j,\vb{y}_j)}{P_\mathrm{KNN}(\bs{\alpha}_\mathrm{nom} | \vb{x}_j,\vb{y}_j)}
    \;.\label{eq:interpolated-ratio}
\end{equation}
Since we do not know the KNN predicted posterior probability for arbitrary $\alpha$, we achieve this by fitting a function for every event that estimates the posterior probabilities for any $\alpha$.
Since $P(\bs{\alpha}_\mathrm{k} | \vb{x}_j,\vb{y}_j)$ are normalized probabilities (over their classes $k$ including the nominal as in \cref{eq:knn-posterior}), the estimated posterior probabilities $\hat{P}(\alpha_k|\vb{x}_j, \vb{y}_j)$ should fulfill the condition that they are always positive and that the probabilities for a specific $\alpha$ sums to one.
To fulfill these constraints, we use polynomial functions of the detector parameters that we pass into the \texttt{softmax} function to get a probability estimate such that
\begin{equation}
\begin{aligned}
    \hat{P}(\alpha_{k}|\vb{x}_j, \vb{y}_j) &= \mathrm{softmax}\left( \vb{g}_j A \right) \\
    &= \frac{\exp(\sum_n g_{jn} A_{nk})}{\sum_{k'} \exp(\sum_n g_{jn} A_{nk'})}
    \;.\label{eq:softmax}
\end{aligned}
\end{equation}
Here, $A_{nk}$ is a matrix of dimension $N \times K$, where $N$ is the number of detector parameters or polynomial features thereof, and $K$ is the number of MC sets.
The entry $A_{nk} = (\alpha_{n,k} - \alpha_{n,\;\mathrm{nom}})^{p_n}$ is the detector parameter difference (dimension $n$) between the $k$th off-nominal and the nominal value. To account for higher order effects it can also be the $p_n$th power thereof.
The vector $\vb{g}_j$ contains the polynomial coefficients (up to chosen order) of this expansion with respect to the detector parameters for the event $j$.
For every event, we minimize the discrete cross-entropy function between the estimated probabilities and the observed ones,
\begin{equation}
    H_j = -\sum_k \log(\hat{P}(\alpha_k|\vb{x}_j, \vb{y}_j)) P_\mathrm{KNN}(\alpha_k|\vb{x}_j, \vb{y}_j)\label{eq:likelihood}\;,
\end{equation}
with respect to the polynomial coefficients $\vb{g}_j$ and store the best fit values for every MC event.
We can then use these fitted polynomial coefficients to calculate an estimate of $r_{j}(\bs{\alpha})$ for any $\bs{\alpha}$, where the normalization in the denominator of \cref{eq:softmax} cancels, simplifying the ratio to
\begin{equation}
    \hat{r}_{j}(\bs{\alpha}) = \frac{\hat{P}(\alpha|\vb{x}_j, \vb{y}_j)}{\hat{P}(\alpha_\mathrm{nom}|\vb{x}_j, \vb{y}_j)} = \exp(\sum_n g_{jn} (\alpha_n - \alpha_{n,\;\mathrm{nom}})^{p_n})
    \;.\label{eq:ultrasurface-weight}
\end{equation}
\cref{fig:ultrasurface-illustration} illustrates the resulting weight function in the space of event properties $\vb{x}$ and detector properties $\alpha$. 
By multiplying all event weights by the ratio  \cref{eq:ultrasurface-weight}, the distribution of nominal MC events can be re-weighted to any point in the space of detector parameters with little computational effort.

%% file: sections/toy_mc.tex
\section{Toy Monte Carlo}
\label{sec:toy-mc}
In order to demonstrate the methods introduced above, we study a simplified neutrino oscillation measurement at a fixed distance and vacuum two-flavor oscillations. We use an analytical description of the hypothesized detector response $P(\vb{y}_j|\vb{x}_j,\bs{\alpha})P(\mathrm{acc}|\vb{x}_j,\bs{\alpha})$, so that the effect of changing $\alpha$ is exactly known for comparison.

\subsection{Sample Generation}

Neutrino oscillations are a phenomenon in elementary particle physics where neutrinos periodically change their flavor as they propagate through space.
An electron neutrino that is produced in a radioactive decay, for example, may be measured by a detector located a few kilometers away from the source as a muon neutrino.
If the detector is only sensitive to the electron flavor, this will manifest as a deficit of measured neutrino interactions compared to the expectation in the absence of oscillations.
The probability that a neutrino produced as flavor $\beta$ is measured as the same flavor, also referred to as its \emph{survival probability}, oscillates as a function of $L/E$, where $L$ is the traveled distance and $E$ is the energy of the neutrino. 
For the purpose of this paper, we assume that the experiment is only sensitive to the survival probability of one flavor that can be approximated as 
\begin{equation}
    P(\nu_\beta\rightarrow\nu_\beta) = 1 - \sin^2(2\theta)\sin^2\left(1.267 \Delta m^2 \frac{L}{E} \Bigg[\frac{\mathrm{km}}{\mathrm{GeV}}\Bigg]\right)
\label{eq:two-flav-oscillation-prob}\;,
\end{equation}
where $\theta$ is known as the \emph{mixing angle} and $\Delta m^2$ is referred to as the \emph{mass splitting}.
The mixing angle determines the amplitude and the mass splitting the frequency of the oscillations.
We assume a fixed distance of \SI{12000}{\kilo\metre} to the source (roughly the length crossing through earth) and values for the mass splitting and mixing angle that are typical for oscillation measurements of neutrinos that are produced in the atmosphere of the Earth ($\Delta m^2 = \SI{2.515e-3}{\electronvolt\squared}$, $\sin^2(\theta)=0.565$), where the oscillation minimum occurs at $\sim \SI{20}{\giga\electronvolt}$ \cite{PDG_2022}.

\begin{figure}[h]
    \centering
    \includegraphics{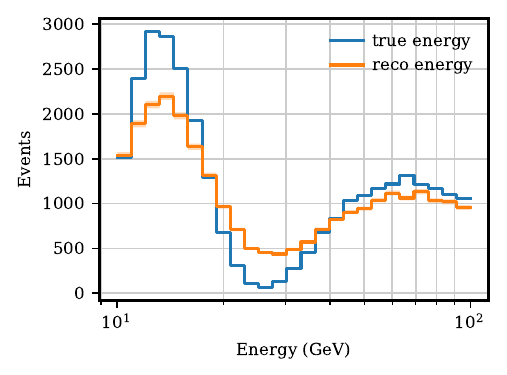}
    \caption{Distribution of true and reconstructed energies from the MC dataset generated with physics parameters $\Delta m^2 = \SI{2.515e-3}{\electronvolt\squared}$ and $\sin^2(\theta)=0.565$ at nominal detector efficiency ($\alpha=1.0$) using \cref{eq:detector_response} (reco).}
    \label{fig:toy-mc-hist-true-reco}
\end{figure}

We sample the true energies of the neutrinos from a log-normal distribution over the range between $\order{\si{\giga\electronvolt}}$ to $\order{\SI{100}{\giga\electronvolt}}$ with its mode at \SI{20}{\giga\electronvolt} to approximately match the energy distribution observed by atmospheric neutrino oscillation experiments \cite{OVS_PRD, super_k}.
After weighting the sampled events with the oscillation probability from \cref{eq:two-flav-oscillation-prob}, the distribution over true energies looks as shown in \cref{fig:toy-mc-hist-true-reco} in blue.
To measure the physics parameters, the detector observes the energy of each detected neutrino and the resulting distribution is binned in a histogram.
The data histogram is then compared to a histogram of simulated neutrino events from MC that is weighted according to \cref{eq:two-flav-oscillation-prob}.

\begin{figure}[h]
    \centering
    \includegraphics{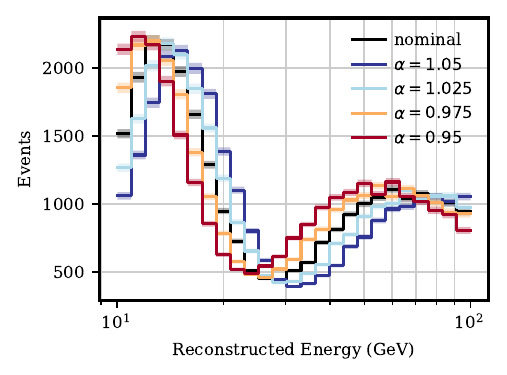}
    \caption{Reconstructed energy distributions for specific choices of the detector parameters, resulting in discrete simulation sets. These would normally be the product of MC simulation, while for illustration purposes we here choose and know an analytic form as in \cref{eq:detector_response}.}
    \label{fig:toy-mc-hist-systematics}
\end{figure}

In our simplified oscillation analysis, the only quantity being measured is the energy and the detector response relates the true energy of the neutrino to the reconstructed energy.
To demonstrate the modeling of detector uncertainties, we assume that the detector has a relative efficiency parameter, $\alpha$, that is only known up to a given uncertainty.
An increase in $\alpha$ causes the reconstructed energy to increase relative to the true energy.
Concretely, the detector response is assumed to follow a normal distribution as a function of the logarithms of the true and reconstructed energies
\begin{equation}
\begin{aligned}
    &P(\log(E_\mathrm{reco}) | \log(E_\mathrm{true}); \alpha, \sigma) \\ 
    &= \mathcal{N}(\mu=(1 + \alpha) \log(E_\mathrm{true}), \sigma)\;.
\end{aligned}
\label{eq:detector_response}
\end{equation}
where an energy smearing in the form of $\sigma$ is included.
For $\alpha=1$ (nominal), the mode of the distribution of the reconstructed energy is centered at the true energy. We assume $\sigma=0.08$ in the following in order to represent a reasonable energy smearing effect. In a more complex scenario, $\sigma$ could also be uncertain to some degree and included as an additional parameter to be treated in the method.
With this smearing, the weighted distribution of reconstructed energies looks as shown in \cref{fig:toy-mc-hist-true-reco} in orange.
As expected, the oscillation minimum is now smeared out compared to the distribution of true energies.
The distributions of the reconstructed energy at different settings of $\alpha$ are shown in \cref{fig:toy-mc-hist-systematics}. 
The location of the oscillation minimum shifts with respect to its true location.

\subsection{Classification and Interpolation}

\begin{figure}[h]
    \centering
    \includegraphics{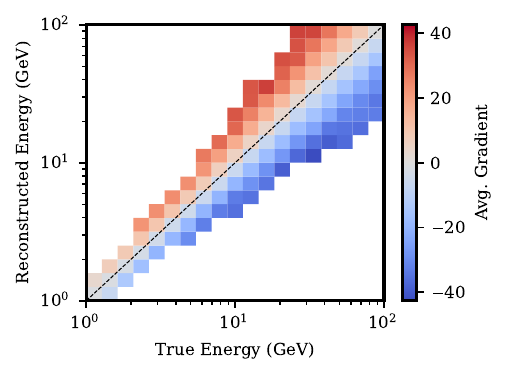}
    \caption{First order polynomial coefficients as a function of true and reconstructed energy. The color in each bin shows the average value of the coefficient with respect to the $\alpha$ parameter in each bin.}
    \label{fig:first-order-gradients}
\end{figure}

Having generated five discrete MC sets ($1e5$ events each) of true and reconstructed energies at different values of $\alpha$, we now apply the KNN classification and interpolation in terms of the polynomial coefficient fitting described in \cref{sec:method-description}.
The inputs into the classifier are the true and reconstructed energy, which we normalize using a power transformer \cite{transformers} before passing them into the KNN, where we set the number of neighbors to 1000 and apply the skew correction described in \cref{sec:linear-tilt-correction}.
Other than this correction, no weighting of the events is used, meaning that no underlying physical model is assumed at this point.
We fit first- and second-order polynomial coefficients with respect to variations in $\alpha$ to the resulting class probabilities of each event as described in \cref{sec:ultrasurface-interpolation}.
The average value of the first order polynomial coefficients as a function of the true and reconstructed energy is shown in \cref{fig:first-order-gradients}.
For events where the reconstructed energy is higher than the true energy, the coefficient is positive, meaning that the event becomes more likely if the detector efficiency is increased.

\begin{figure}[h]
    \centering
    \includegraphics{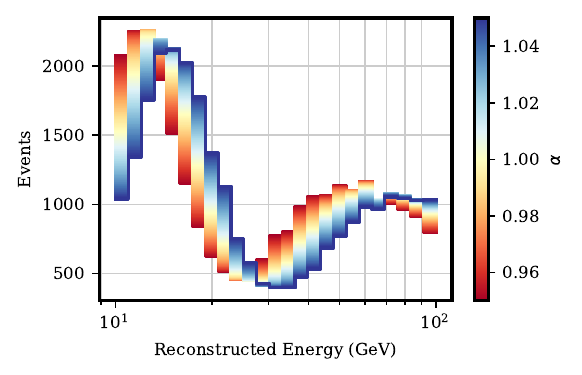}
    \caption{Predicted event counts as a result of sweeping over the $\alpha$ parameter and weighting the nominal MC events with $w(\alpha$), which can now be done in a continuous manner, by applying the new method.}
    \label{fig:histogram-parameter-sweep}
\end{figure}

\begin{figure}[h]
    \centering
    \includegraphics{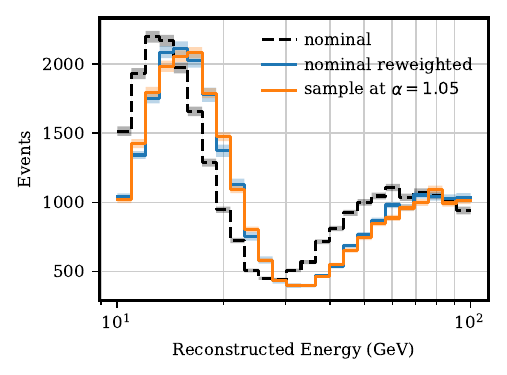}
    \caption{Comparison between the weighted distributions of the nominal MC set, a set sampled independently at $\alpha = 1.05$ and the nominal MC weighting by $w(\alpha=1.05)$.}
    \label{fig:ultrasurface-reweighting-distribution}
\end{figure}

Using all fitted polynomial coefficients, we can now transform the nominal distribution smoothly to that at any point in $\alpha$ within the range covered by the systematic sets.
This is shown in \cref{fig:histogram-parameter-sweep}, in which we sweep over the $\alpha$ parameter.
In \cref{fig:ultrasurface-reweighting-distribution}, we show the result of weighting the nominal MC set by $w(\alpha=1.05)$ compared to a new sample that was generated independently at the same value of $\alpha$.
The agreement between the two lies well within the statistical uncertainties.
Most importantly, this agreement is achieved in the distribution of MC events after they have been weighted with the neutrino oscillation weights from \cref{eq:two-flav-oscillation-prob}, despite the fact that the polynomial coefficients were fit on unweighted MC events.
This demonstrates that the event-wise polynomial coefficients found using our method are indeed fully decoupled from the assumed physics parameters, which is the main goal of this work.

\subsection{Performance}

For any given point in $\alpha$, we evaluate the performance of our method by comparing the re-weighted nominal MC set to an independently generated MC set at that value of $\alpha$. The method works if the $\chi^2$ between the corresponding histograms is compatible with statistical fluctuations.

\subsubsection{Hyper-parameter Optimization}
\begin{figure}[h]
    \centering
    \includegraphics{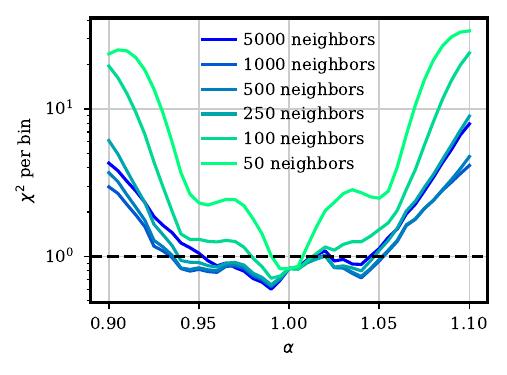}
    \caption{Comparison of histograms from the re-weighted nominal MC set and independently generated MC sets at different values of $\alpha$. Shown is the $\chi^2$ test statistic per number of bins as a function of $\alpha$ for different settings of the number of neighbors used in the KNN classifier. The range covered by the discrete simulation sets is only from 0.95 to 1.05.}
    \label{fig:chi2-vs-n-neighbors}
\end{figure}

We compare different settings of the number of neighbors used by the KNN classifier and catch both underfitting and overfitting failure modes. 
As described earlier in \cref{sec:simple-knn}, the classification output will be dominated by random fluctuations if the number of neighbors is too small, leading to overfitting.
As a result, the re-weighted nominal set will match the distribution of the MC set that went into the fit exactly, but its match with an independently drawn MC set will be poor.
This effect can be observed in \cref{fig:chi2-vs-n-neighbors}, where we show the $\chi^2$ test statistic as a function of $\alpha$ for different numbers of neighbors used in the KNN. 
At $\alpha_\mathrm{nom}=1$, all weights generated by \cref{eq:ultrasurface-weight} are equal to one by construction and the bin-wise $\chi^2$ is only due to statistical fluctuations.
Moving away from the nominal point, the agreement becomes poor when only 50 or 100 neighbors are used, because the fitted polynomial coefficients are mostly representing statistical fluctuations and do not model the actual systematic effects.
The polynomial coefficients that are fit to the KNN output using 250 to 1000 neighbors perform similarly very well within the range $[0.95, 1.05]$  that is spanned by the systematic sets, showing values of $\chi^2$ that are compatible with statistical fluctuations alone.
Extrapolation beyond this range, however, does not work very well for any setting of neighbors, and we therefore recommend that MC sets should always be produced over the entire range of nuisance parameters that is of interest for an analysis. This is also necessary in order to include all effects of bin migration of the histograms used in the analysis later. Our method includes these naturally, since it is employed at the event level.
Some underfitting is only observed at a setting of $N=5000$, which corresponds to 10\% of the entire dataset including the nominal and off-nominal samples.

\subsubsection{Decoupling from the Physics Model}

\begin{figure}[h]
    \centering
    \includegraphics{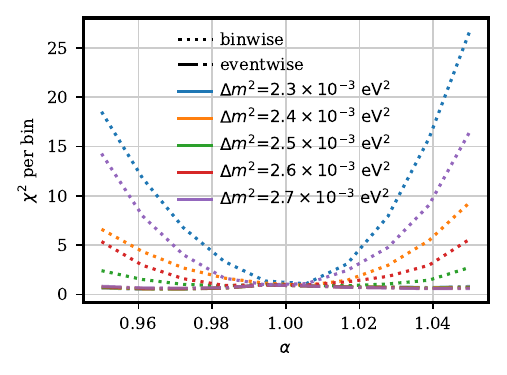}
    \caption{Comparison of histograms from the re-weighted nominal MC set and independently generated MC sets at different values of $\alpha$ and $\Delta m^2$. Shown is the $\chi^2$ test statistic per number of bins as a function of $\alpha$ for different mass splitting values $\Delta m^2$. The dashed-dotted line shows the result of using the re-weighting method presented in this manuscript. We obtain a good agreement over the full phase space considered. For comparison, the dotted line shows the result of applying binwise corrections, which deviates for large differences in $\Delta m^2$.}
    \label{fig:chi2-vs-mass-splitting}
\end{figure}

Similarly, we test the predicted event counts from our re-weighting method, but now targeting different physics parameters. We apply the ratio of \cref{eq:ultrasurface-weight} to correct for variations in the detector systematic parameter value $\alpha$, and also apply $w_\mathrm{flux}=\Phi(\vb{x}|\bs{\theta})/\Phi_{\mathrm{sim}}(\vb{x})$ when predicting different mass splitting values $\Delta m^2$, for a fixed value of $\sin^2(\theta)=0.565$, see \cref{eq:two-flav-oscillation-prob}. We compare to an independently generated MC set at target values of $\alpha$ and $\Delta m^2$, and quantify the agreement in terms of the $\chi^2$ statistic as shown in \cref{fig:chi2-vs-mass-splitting}. We see that for the full phase space considered here, we obtain an excellent performance for the \textit{eventwise} method. For comparison, we demonstrate the application of \textit{binwise} gradients $\grad_\alpha \mu_i$ calculated in every analysis bin, see \cref{sec:problem-formulation}. This latter method needs to assume specific physics parameter values ($\Delta m^2=2.515 \times 10^{-3} \mathrm{eV}^2$ in this example) when calculating these gradients and performs badly if variations in both $\alpha$ and $\Delta m^2$ are to be considered.
This demonstrates how our method of taking detector systematic parameters into account is independent of the physics parameters to be measured.

\subsection{Possible Extensions to the Method}
\label{sec:extensions}
In this paper, we demonstrated the application of our method to the simple case of only one detector parameter.
In principle, however, it can be applied to any number of parameters and the polynomial form of the weight functions allows for any order of polynomial coefficients and even correlation terms between parameters.
We also presented the case that the nominal MC set and the systematic MC sets are of similar statistical power, yet we only use the nominal MC set to produce predictions.
This is not the most efficient way to use the available MC simulation, because the statistical uncertainty on the prediction is similar to that of a single MC set.
However, there is nothing special in principle about the nominal set. 
The weight function defined in \cref{eq:ultrasurface-weight} can re-weight events from any systematic set to any value of $\alpha$. 
It is therefore in principle possible to use the simulated events of all MC sets together, which should greatly increase the  statistical power of the prediction and make more efficient use of limited computational resources.

In \cite{2019_Snowstorm}, it was shown that there are advantages to produce a MC set with continuously varied detector properties, rather than several MC sets for discrete points in the parameter space.
Such an approach is even more efficient with the available computational resources, especially when the number of detector parameters is large.
Although we demonstrated our method only for the case of discrete MC sets, the mathematical formulation in principle also allows the use of a continuously varied MC set. 
Effectively, this would mean treating each event as representing its own class with only one member with its individual value of the detector parameters, where the posterior probability is equal to $1/N$ if it is inside the neighborhood around a point and zero otherwise.
The loss function from \cref{eq:likelihood} that is minimized to fit the polynomial coefficients would be
\begin{equation}
    \mathcal{L} = -\sum_{k\in\mathcal{N}(x)} \log(q(\alpha_k))\;,
\end{equation}
where $\mathcal{N}(x)$ is the set of indices that belong to the neighborhood around the query point $x$.
The only difference in the application of the weights in this case would be that the value of $\alpha_\mathrm{nom}$ would be different for each event in the sample.
The detailed testing of this method and the comparison between the performance of the discrete MC case presented in this work and the case of continuously sampled MC sets is beyond the scope of this paper. It promises to be an interesting avenue for exploration in future work, though, since our novel method does not require linearity of the observable distribution with varying detector response parameters $\alpha$, which is an assumption made in the original work \cite{2019_Snowstorm}.

%% file: sections/conclusion.tex
\section{Conclusion} \label{sec:conclusion}

\subsection{Summary}

In this work, we have demonstrated a novel method of event-wise interpolation in a large set of MC simulated events. This allows to re-weight these events to model the effects of variations in the properties of a particle physics experiment in the context of a MC forward-folding analysis.
This event-wise interpolation effectively models the variations in the detector response in a way that is fully decoupled from assumptions about the physics parameters that are being measured.
To achieve this, we use a simple, non-parametric KNN classifier to calculate the posterior probability that a given event belongs to any of the MC sets that are produced under varied realizations of the detector properties.
We then fit polynomial coefficients to the posterior distributions of each event by minimizing the negative log-likelihood between the observed and predicted probabilities.
With this method, we are able to model the effects of systematic uncertainties for every event up to any polynomial order in all parameters that describe the detector properties, including correlation terms between these parameters.
Using a toy MC example that is inspired by fixed distance neutrino oscillation experiments, we demonstrated the performance of our method in the case where the detector only has one parameter of uncertainty, $\alpha$, which resembles something similar to an energy scale systematic uncertainty.
We could show that, even though the polynomial coefficients were fit on \emph{unweighted} MC events, we can re-weight the nominal MC set at $\alpha=1$ to any value of $\alpha$ within the range spanned by the discrete off-nominal MC sets and match the \emph{weighted} distributions to within errors from statistical fluctuations.
Our method therefore solves the task of fully decoupling the modeling of the detector response from the weights that are associated with the physics parameters one wishes to measure in a forward-folding analysis. 

\subsection{Outlook}
The problem of modeling particle detector systematic uncertainties when there is no analytical form of the effect of these uncertainties is common in particle physics and several methods have been proposed in the past to solve it.
Most of these methods, however, attempt to model detector effects only in the space of the reconstructed variables that are used to perform the measurement. 
We showed in \cref{sec:problem-formulation} that such an approach necessarily leads to a coupling between the detector response and the physics parameters. 
Our method, in contrast, extracts the detector response as a function of both true and reconstructed variables in a way that decouples it fully from event weights that are functions of the physics parameters. 

The classification method used here is a simple, deterministic calculation with only one tunable hyperparameter. In the development of this work, we found that more sophisticated classification methods, such as neural networks, could only approach the performance of the KNN after many iterations of hyperparameter tuning and never showed a clear advantage over it. We emphasize that this will eventually depend on the exact use-case, with specific detector response uncertainties and limits to the available MC statistics that will dictate the optimal classified for the problem.

Since the novel method presented here produces a re-weighting scheme on the event level, rather than the histogram level, the observable space can be rebinned without any need to re-compute the effect of detector response parameter variations. This makes analysis binning optimization studies very efficient and could be helpful for approaches like \cite{DeCastro:2018psv}. As an unbinned method taking into account systematic uncertainties in the detector response, our method also paves the way for the use of unbinned likelihood formulations in MC forward-folding analyses.


%% file: sections/acknowledgements.tex
\section*{Code}
The code to reproduce the MC simulations and figures shown in this paper can be found at \url{https://github.com/LeanderFischer/ultrasurfaces}.

\section*{Acknowledgments}

The authors acknowledge support from DESY (Zeuthen, Germany), a member of the Helmholtz Association HGF, and thank Michelle Tsirou, Summer Blot, and Sofia Athanasiadou for reviewing the paper and providing valuable feedback.

%% file: sections/supporting.tex
\section{K-Nearest Neighbors - Linear Skew Correction}
\label{sec:linear-tilt-correction}

While the "naive" KNN performs reasonably well in densely sampled regions of the parameter space, it suffers from edge effects when reaching the tails of the input distributions. The reason for this is that the calculation assumes that the samples are distributed uniformly within the neighborhood of the queried point in $\vb{x}$. However, towards the tails of the distribution or in regions with rapid changes in the density, this assumption is broken. For a query point on the right side of the normal distribution in our example, most of the counted neighbors are going to be located to the left of the queried point. As a result of this skew in the sample distribution, the \emph{effective} query point at which the ratio is calculated is biased towards the left.

A simple way to correct for this skew within a neighborhood of samples is to weight them in such a way that their \textit{center of gravity} (COG) coincides with the point being queried.
We achieve this by assuming a linear relationship between the weight and the position of each sample with a gradient $u_i$ that is independent for each dimension. The weight of each sample is then
\begin{equation}
    w_{j} = 1 + \sum_i u_{i} \Delta x_{ij}\;,\label{eq:tilt-weights-linear}
\end{equation}
where $u_i$ is the gradient of the weight with respect to the $i$th dimension and $\Delta x_{ij}$ is the distance between the $j$th sample and the query point in the same dimension.
We calculate the gradients for each dimension and class independently by setting the COG in that dimension to zero and solving
\begin{align}
    \sum_{j\in \mathcal{N}_k(x)} (1 + u_{i} \Delta x_{ij}) \Delta x_{ij} &= 0 \\
    \Leftrightarrow u_i &= \frac{\sum_{j\in \mathcal{N}_k(x)} - \Delta x_{ij}}{\sum_{j\in \mathcal{N}_k(x)} (\Delta x_{ij})^2}
    \;,
\end{align}
where $\mathcal{N}_k(x)$ is the set of indices of the $N$ nearest neighbors of $x$ that belong to class $k$. While a correction with these gradients correctly removes the skew of an approximately uniform distribution, it can produce invalid negative weights when the difference between the query point and the average sample position is large, as is the case in regions far outside of the input distributions. 
To fix this, we evaluate the weights using an exponential function such that
\begin{equation}
    w_{j} = \prod_i \exp(u_i \Delta x_{ij})\;.
\end{equation}
This approximates \cref{eq:tilt-weights-linear} in small, nearly uniform neighborhood regions and produces small but non-zero weights for highly skewed ones.

\section{Detailed Derivation of Posterior Ratio}
\label{sec:posterior-ratio-derivation}

In this section, we derive the equality between the ratio of detector response probability functions and the classification posterior ratio in \cref{eq:ratio-posterior}.
For the purpose of the derivation, we drop the event index $j$ from the equation. 
We start by noting that the detector response PDF is conditional on an event having been accepted, that is, $P(\vb{y}|\vb{x},\bs{\alpha}) \equiv P(\vb{y}|\vb{x},\bs{\alpha}, \mathrm{acc})$.
We then rearrange the conditional probabilities as follows, where we apply Bayes' theorem where appropriate
\begin{equation}
    \begin{aligned}
        P(\vb{y}|\vb{x},\bs{\alpha}, \mathrm{acc}) P(\mathrm{acc}|\vb{x},\bs{\alpha})
        &=
        \frac{
            P(\vb{x},\vb{y},\bs{\alpha},\mathrm{acc})
        }{
            P(\vb{x},\bs{\alpha})
        }\\
        &=
        \frac{
            P(\bs{\alpha}|\vb{x},\vb{y},\mathrm{acc}) P(\vb{x},\vb{y},\mathrm{acc})
        }{
            P(\vb{x}|\bs{\alpha})P(\bs{\alpha})
        }
        \;.
    \end{aligned}
\end{equation}
Therefore, the ratio from \cref{eq:ratio-posterior} becomes
\begin{equation}
    \frac{P(\vb{y}|\vb{x},\bs{\alpha}_k)P(\mathrm{acc}|\vb{x},\bs{\alpha})}
     {P(\vb{y}|\vb{x},\bs{\alpha}_\mathrm{nom})P(\mathrm{acc}|\vb{x},\bs{\alpha}_\mathrm{nom})}
     =
     \frac{
        P(\bs{\alpha}_k|\vb{x},\vb{y},\mathrm{acc}) 
    }{
        P(\vb{x}|\bs{\alpha}_k)P(\bs{\alpha}_k)
    }
    \frac{
        P(\vb{x}|\bs{\alpha}_\mathrm{nom})P(\bs{\alpha}_\mathrm{nom})
    }{
        P(\bs{\alpha}_\mathrm{nom}|\vb{x},\vb{y},\mathrm{acc}) 
    }\label{eq:ratio-derivation-1}
    \;.
\end{equation}
If we assume that the \emph{true} event properties $\vb{x}$ have been sampled in the same way for every systematic set, then $P(\vb{x}|\bs{\alpha}_\mathrm{nom}) = P(\vb{x}|\bs{\alpha}_k)$. If we furthermore assume that the same number of events have been originally sampled for every set (that is, \emph{before} the application of the detector response), then $P(\bs{\alpha}_\mathrm{nom}) = P(\bs{\alpha}_k)$ and everything except the ratio of posteriors cancels in \cref{eq:ratio-derivation-1}. Finally, we can drop the condition that an event was accepted by the detector because it is always true for every event for which the reconstructed quantities $\vb{y}$ exist, meaning that $P(\bs{\alpha}|\vb{x},\vb{y},\mathrm{acc}) \equiv P(\bs{\alpha}|\vb{x},\vb{y})$. We thus arrive at the expression in \cref{eq:ratio-posterior}. In some analysis scenarios, it may be useful to generate different numbers of events per set, or to even sample the true event parameters differently in each set. In these cases, one can reintroduce the factors $P(\bs{\alpha})$ and $P(\vb{x}|\bs{\alpha})$ to weight events in such a way as to compensate for their relative over- or under-representation in the input sets.